# Secure Combination of Untrusted Time information Based on Optimized Dempster-Shafer Theory

Yang Li, Yujie Luo, Yichen Zhang, Ao Sun, Wei Huang, Shuai Zhang, Tao Zhang, Chuang Zhou, Li Ma, Jie Yang, Mei Wu, Heng Wang, Yan Pan, Yun Shao, Xing Chen, Ziyang Chen, Song Yu, Hong Guo*, and Bingjie Xu*

*Abstract*—Secure precision time synchronization is important for applications of Cyber-Physical Systems. However, several attacks, especially the Time Delay Attack (TDA), deteriorates the performance of time synchronization system seriously. Multiple paths scheme is thought as an effective security countermeasure to decrease the influence of TDA. However, the effective secure combination algorithm is still missed for precision time synchronization. In this paper, a secure combination algorithm based on Dempster-Shafer theory is proposed for multiple paths method. Special optimizations are done for the combination algorithm to solve the potential problems due to untrusted evidence. Theoretical simulation shows that the proposed algorithm works much better than Fault Tolerant Algorithm (FTA) and the attack detection method based on single path. And experimental demonstration proves the feasibility and superiority of the proposed algorithm, where the time stability with 27.97 ps, 1.57 ps, and 1.12 ps at average time 1s, 10s, 100s is achieved under TDA and local clock jump. The proposed algorithm can be used to improve the security and resilience of many importance synchronization protocol, such as NTP, PTP, and TWFTT.

## I. INTRODUCTION

With the development of industrial internet of things, the applications of Cyber-Physical Systems (CPS) such as autonomous vehicle, autonomous mobile robots, automated factory and smart grid are becoming increasingly popular. In order to achieve synchronous communication and computation on different spatial and temporal scales for CPS, high precise time synchronization is one of the most important requirements.

One the one hand, different time synchronization precision from millisecond level to sub-nanoseconds level are required for different applications. There are already many time synchronization technologies, such as Network Time Protocol (NTP) [1], Precision Time Protocol (PTP)[2], and Two-Way Fiber-Optic Time Transfer (TWFTT) [3], and Global Navigation Satellite System (GNSS), which achieve different time synchronization precisions. On the other hand, security is extremely important for CPS and attacks on time synchronization can lead to disastrous effects. For example, damages may happened to the smart grid by the wrong operation due to deteriorated time synchronization [4].

In [5], several attacks are analyzed, such as packet manipulation, spoofing, replay attack, and a set of security requirement is defined for time protocols in packet switched networks. Since information data and physical signal are simultaneously used for time synchronization, the complete requirements of secure time synchronization should include countermeasures of attacks for the two parts. The cryptography method is an efficient method to protect the information data from malicious manipulation [6]. However, an attack called time delay attack (TDA), which changes the flying time of the time signal, is the frailty of time synchronization and deteriorates the performance seriously. Many methods are proposed to decrease the influence of TDA [7-14]. The first kind of methods is to utilize the historical data from the single link of time synchronization, such as the delay attack detection and mitigation methods for PTP [7] and TWFTT [8]. The second kind of methods is to utilize extra information, such as information from phase measurement unit (PMU) [9] or external reference clock [10]. The third

This work was supported in part by the Equipment Advance Research Field Foundation (Grant number 315067206), the National Key Research and Development Program of China (Grant No. 2020YFA0309704), the National Natural Science Foundation of China (Grant Nos U19A2076, 62101516, 62171418, 62201530), the Sichuan Science and Technology Program (Grant Nos 2022ZYD0118, 2023JDRC0017, 2023YFG0143, 2022YFG0330, 2022ZDZX0009 and 2021YJ0313), the Natural Science Foundation of Sichuan Province (Grant Nos 2023NSFSC1387 and 2023NSFSC0449), the Basic Research Program of China(Grant No. JCKY2021210B059), the Equipment Advance Research Field Foundation(Grant No. 315067206), the Chengdu Major Science and Technology Innovation Program (Grant No. 2021-YF08-00040-GX), the Chengdu Key Research and Development Support Program (Grant Nos 2021-YF05-02430-GX and 2021-YF09-00116-GX), the Foundation of Science and Technology on Communication Security Laboratory (Grant No. 61421030402012111). Corresponding authors: xbjpku@pku.edu.cn, hongguo@pku.edu.cn.

Yang Li, Yujie Luo, Ao Sun, Wei Huang, Shuai Zhang, Tao Zhang, Chuang Zhou, Li Ma, Jie Yang, Mei Wu, Heng Wang, Yan Pan, Yun Shao, and Bingjie Xu are with National Key Laboratory of Security Communication, Institute of Southwestern Communication, Chengdu 610041, China.

Yichen Zhang, Xing Chen, and Song Yu are with State Key Laboratory of Information Photonics and Optical Communications, Beijing University of Posts and Telecommunications, Beijing 100876, China.

Ziyang Chen and Hong Guo are with State Key Laboratory of Advanced Optical Communication Systems and Networks, School of Electronics, and Center for Quantum Information Technology, Peking University, Beijing 100871, China.

important kind of methods is to utilize multiple paths or multiple clocks. This method can detect the delay attack, and improve the resilience of the system [11-13]. Many studies have been done for multiple paths or multiple clocks methods. In [1], a complex convergence algorithm including selection, clustering and combining is proposed for NTP. However, this method is hard to be applied to more precise time synchronization protocol. In [14], a method based on Kalman filter is proposed for the multiple paths synchronization. It does not focus on detecting TDA, but to decrease the influence of TDA. So, its efficiency becomes poor when TDAs happened on more than one path.

Based on those studies, the method of multiple paths or multiple clocks is added in IEEE-1588-2019 standard [2] as one of the four security features. It is recommended in the standard that the suspected time source or path may be removed by a voting algorithm. However, no voting algorithm is recommended in the standard. In [15], the Fault Tolerant Algorithm (FTA) is proposed to mitigate the path failures. In [16], the FTA is extended to solve the fail-silent clocks problem. Although the FTA is shown to be a useful combination method, it still has some limitations. Firstly, since only the maximum and minimum offset is dropped, if more malicious paths are suspected, the offsets from the undetected malicious paths deteriorate the performance. Second, if there are no suspected paths, the dropped information may lead to reduced performance.

In this paper, we proposed a combination algorithm based on a decision-level data fusion method called Dempster-Shafer (D-S) theory [17, 18]. In fact, the D-S theory is a robust and flexible mathematical tool for modeling and merging uncertain, imprecise, and incomplete data, and is widely used in multisensory data fusion applications such as fault diagnosis, pattern recognition and so on. Due to the potential delay attacks in arbitrary unknown paths, the time information is uncertain. By using D-S theory, the malicious paths can be detected, and the time information from the health paths is used to update the slave clock. In this paper, the optimization of the combination method based on D-S theory is studied to manage the conflicting evidences due to untrusted time information. The simulation results show that the optimized combination method works better than FTA method and attack detection method based on time information of single path no matter whether there are TDAs or whether there is slave clock jump. Experimental demonstration on a 3 paths synchronization system is implemented to prove the feasibility and priority of the proposed method, that the time stability with 27.97 ps, 1.57 ps, and 1.12 ps at average time 1s, 10s, 100s is achieved under TDA and local clock jump.

## II. SCHEMATIC DESCRIPTION

*A. Multiple Paths Scheme and Dynamic Equations*

We consider a general scheme of the time synchronization system with multiple paths (as shown in Fig. 1). It consists of two parties, master site and slave site. There are N paths between the two parties. For each path, there are a master port in the master site and a slave port in the slave site, which are interconnected by a channel, for example, optical fiber. For each path, a kind of synchronization protocol is applied to obtain the time offset between the master clock and the local clock, such as PTP, NTP, or TWFTT. In fact, the proposed method can also be applied for the multiple master clocks problem. For simplify, we only focus on the multiple paths problem in this paper.

In order to synchronize with the remote clock, the slave clock is updated with time offset correction, $u_\theta(t_n)$, periodically. The synchronization instant is given as $t_n = n \cdot \tau$. Without loss of generality, $\tau$ is set to be 1 second in this paper.

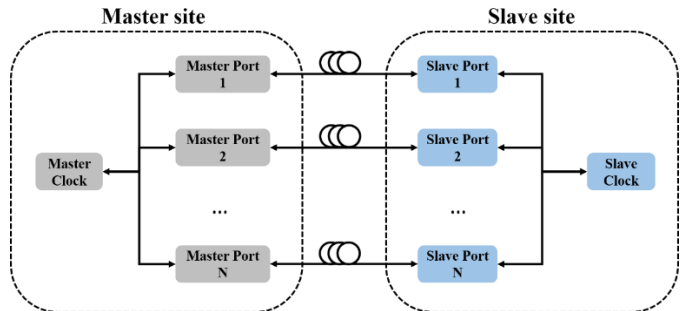

**Fig. 1.** Schematic diagram of multiple paths time synchronization system.

By employing a two-state clock model, the equations for the clock dynamics are [8]

$$\begin{cases} \theta(t_n) = \theta(t_{n-1}) + u_\theta(t_{n-1}) + \gamma(t_{n-1}) \cdot \tau + \omega_\theta(t_n) \\ \gamma(t_n) = \gamma(t_{n-1}) + \omega_\gamma(t_n) \end{cases}, (1)$$

where $\theta(t_n)$ and $\gamma(t_n)$ are time offset and frequency difference, $\omega_\theta(t_n)$ and $\omega_\gamma(t_n)$ are independent Wiener processes with zero-mean and variances equal to $\sigma_\theta^2$ and $\sigma_\gamma^2$, and $u_\theta(t_{n-1})$ is the update value of the slave clock at $t_{n-1}$.

For each path, the measured time offset from the i'th path, $\theta_M^i(t_n)$, is given by

$$\theta_M^i(t_n) = \theta(t_n) + \omega_d^i(t_n) + \omega_m^i(t_n) + \tau_{attack}^i(t_n), \quad (2)$$

where $\omega_d^i(t_n)$ and $\omega_m^i(t_n)$ are transmission and measurement noises for the i'th path, which are modeled as independent Wiener processes with zero-mean and variances equal to $\sigma_{d,i}^2$ and $\sigma_{m,i}^2$, and $\tau_{attack}^i(t_n)$ is time error caused by the possible TDA in the i'th path. If there is no TDA, $\tau_{attack}^i(t_n) = 0$. The bias which is caused by the path asymmetry for two-way synchronization or path uncertainty caused by the one-way synchronization is ignored for simplify.

*B. D-S Evidence Theory*

Since the D-S theory is an effective method to merging uncertain information, so it is widely used in multisensory information fusion applications, such as fault diagnosis. For multiple path time synchronization system, each path can be treated as a special sensor. The influence of TDA on one path in fact brings fault result on the sensor information. So, the D-S theory can be used to detect the TDA for multiple path time synchronization system.

For the D-S theory, the first fundamental concept is the frame of discernment, which is the set of all possible states, $\Theta = \{\theta_1, \theta_2, \ldots, \theta_n\}$. All subsets of $\Theta$ is named the hypothesis set and denoted as $2^\Theta$.

The second fundamental concept is the basic probability assignment (BPA) or mass function, which is defined as the probability function for each hypothesis. The BPA is defined as

$$m: 2^\Theta \to [0,1],$$

where

$$m(\emptyset) = 0,$$
$$m(H) \geq 0, \forall H \subset \Theta,$$
$$\sum_{H \subset \Theta} m(H) = 1.$$

Suppose $m_1$ and $m_1$ are BPA over the same frame $\Theta$, the D-S theory defines the rule of combination as

$$m_1 \oplus m_2(\emptyset) = 0, \quad (3)$$
$$m_1 \oplus m_2(H) = \frac{\sum_{i,j,X_i \cap Y_j = H} m_1(X_i) m_2(Y_j)}{1 - \sum_{i,j,X_i \cap Y_j = \emptyset} m_1(X_i) m_2(Y_j)}, \quad (4)$$

From the definition, it can be seen that a priori probabilities are not required in the D-S theory. So, it is suitable to detect the attacks unseen previously.

*C. Application of D-S Evidence Theory to Secure Clock Combination Problem*

For each time synchronization path, D-S evidence theory is applied to distinguish whether a TDA happens to the path. From section II. B, the frame of discernment and explicit expression of BPS should be given.

In this paper, the frame of discernment which consists of two elements, normal and attack, is adapted for secure clock combination problem.

The BPAs is the probability function for each hypothesis. For time synchronization, the measured time offset is a useful variable for attack detection, such as in [8]. However, for each path of time synchronization, there are only one measured time offset since there is only one slave clock. Suppose $\theta_M^i$ is the measured time offset from the i'th path, then $\theta_M^i - \theta_M^j$ can be used as a virtual measured time offset by treating the j'th path as a virtual slave clock. So, if there are N paths, there are N measured time offsets which can be used for attack detection. In order to apply the D-S evidence theory, it is needed to transfer the measured time offsets to BPAs.

In this paper, a modified sigmoid function is chosen as the BPA.

$$m(attack) = \frac{1}{1 + e^{-A(x-B)}}, \quad (6)$$
$$m(normal) = 1 - m(attack), \quad (7)$$

where $x$ is the absolute value of measured time offset or virtual measured time offset. In order to improve the detection precision, the estimated frequency difference is subtracted from the measured time offset, such as in [8]. The parameters A and B are positive. So, m(attack) approaches to 1 when $x$ approaches to infinity. The parameter A is an empirical data which determine the speed at which m(attack) approaches to 1 when $x$ is larger than B. The value of B is the critical value that when $x$ equals to B the probabilities of attack and normal are the same. In this paper, the value of B is given as follows. Similar to [19] and [20], the allowed false alarm rate is $p_F$ and the allowed missed detection rate is $p_M$, as shown in Fig. 2.

Firstly, the threshold T is given by solving the equation (8)

$$p_F = 1 - \int_{-\infty}^{T} f(\theta) d\theta, \quad (8)$$

where $f(\theta)$ is the probability density function when there is no attack.

Secondly, the minimum detectable delay attack L is given by solving the equation (9)

$$p_F = \int_{-\infty}^{T} f_L(\theta) d\theta, \quad (9)$$

where $f_L(\theta)$ is the probability density function when the influence of delay attack on the measured time offset is L.

Finally, the value of B can be given by the intersect point of $f(\theta)$ and $f_L(\theta)$.

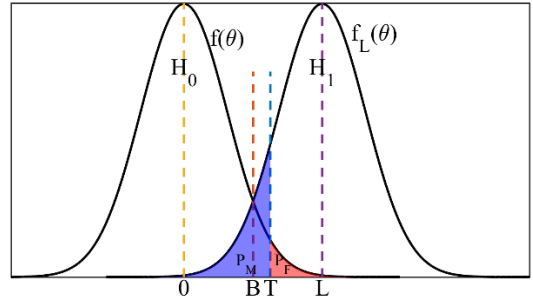

**Fig. 2.** Representation of the probability density function with the threshold T, minimum detectable delay attack L, false alarm $p_F$, and missed detection $p_M$.

*D. Metric*

In order to analyze the effect of the algorithm proposed in this paper, two kinds of metrics are introduced. The first kind is precision and recall, The definitions of precision and recall are Precision $= TP/(TP + FP)$ and Recall $= TP/(TP + FN)$, where TP is the number of detected actual attack events, FP is the number of detected wrong attack events, and FN is the number of undetected actual attack events. The second kind is time deviation error variance (TDEV), which follows the definition in [21]

$$\text{TDEV}(\tau = n*\tau_0) = \sqrt{\frac{1}{6n^2(N-3n+1)} \sum_{j=0}^{N-3n} \left[\sum_{i=j}^{n+j-1}(x_{i+2n} - 2x_{i+n} + x_i)\right]^2} \cdot \quad [10]$$

### III. SIMULATION AND ANALYSIS

*A. Optimization of Secure Clock Combination Method*

For D-S theory, management of conflicting evidences is very important [22]. Unreasonable results may arise when conflicting evidences are combined. For example, as shown in section II.A, when the time offset from the j'th path is used to distinguish whether the i'th (i ≠ j) path is a malicious path, $\theta_M^i - \theta_M^j$ is used. If the i'th path is not a malicious path and the j'th path is a malicious path, the unknown delay in $\theta_M^j$ caused by the attack in the j'th path brings wrong information for the attack detection. Since there is no priori knowledge on whether j'th path is malicious, $\theta_M^i - \theta_M^j$ is an untrusted evidence. Although the final decision depends on the information from all the paths, if the delay caused by the attack is very large, the original clock combination method based on D-S theory may bring wrong result. This problem is called Zadeh's paradox [23]. Two kinds of methods are proposed to solve this problem [22]. In this paper, we solve the problem by limiting the maximum value of the BPA.

$$m(attack) = \min\left[\frac{1}{1 + e^{-A(x-B)}}, k_{max}\right].$$

Actually, the value of $k_{max}$ is a measure of credibility. Since the evidence is untrusted, the probability derived from it should be limited.

However, it is shown blow that the limitation of $m(attack)$ also brings new problems. Since $m(normal) = 1 - m(attack)$, the maximum limitation of $m(attack)$ increases the minimum value of $m(normal)$. That means if the i'th path is a malicious path, the probability that it is normal derived from the evidence is increased due to the limitation. If the j'th path is also a malicious path with the same time error caused by delay attack, the evidence from the j'th path makes $m(normal)$ approach 1. Then, even though more evidences show that the i'th path is a malicious path, the final decision may be that it is a normal path. Then the delay attack is not detected. The problem is indeed that the evidence that the path is normal may also be untrusted when the two paths are both malicious paths.

In all, if the j'th path is a malicious path, the evidence from the j'th path may be untrusted when the i'th path is a normal path and the evidence from the j'th path may also be untrusted when the i'th path is a malicious path. The first problem can be solve by limiting the maximum value of $m(attack)$. The second problem can also be solve by limiting the maximum value of $m(normal)$, or, by limiting the minimum value of $m(attack)$.

$$m(attack) = \max\left[\frac{1}{1+e^{-A(x-B)}}, k_{min}\right].$$

In the following, two simulations are designed to show the importance of limiting the maximum value and minimum value of $m(attack)$.

Three algorithms are compared in the simulations. The first algorithm (DS0) only adopts the original definition of $m(attack)$. The second algorithm (DS1) adopts the original definition of $m(attack)$ and the maximum limitation. The third algorithm (DS2) adopts the original definition of $m(attack)$, the maximum limitation, and the minimum limitation.

The case of five paths synchronization system is simulated. Without loss of generality, the variances of transmission and measurement noises are the same for each path, respectively. The values of simulation parameters are shown in Table.1.

TABLE I
SIMULATION PARAMETERs

| Parameter | Value | Description |
|---|---|---|
| $\sigma_{m,i}$ | 25 ps | Standard deviation of measurement noise |
| $\sigma_{d,i}$ | 10 ps | Standard deviation of transmission noise |
| $\sigma_\theta$ | 10 ps | Standard deviation of random-walk phase noise of the clock |
| $\sigma_\gamma$ | 1 ps/s | Standard deviation of random-walk frequency noise of the clock |

The results of the first simulation is shown in Fig. 3. In the simulation, TDA happens once every 50s, and only one path is attacked each time. For the i'th (i = 1,2,3,4,5) path, TDA happens when $\mod(t_n - (i-1)*10s, 50s) == 0$, where $\mod(x)$ is the remainder function. In order to show the influence of large TDA on the original algorithm, the value of TDA is set to be 10ns. For the simulation, 40 TDAs are applied to each path.

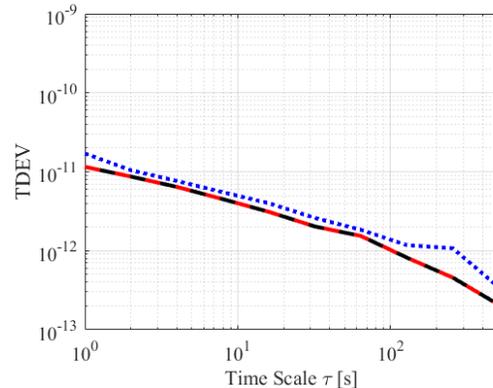

**Fig. 3.** Simulation of 5 paths time synchronization system's TDEV with one path under TDA once every 50s. (dot line(blue): DS0; dash line(red): DS1; solid line(black): DS2).

TABLE II
PERFORMANCE METRIC WITH ONE PATH UNDER TDA ONCE EVERY 50s

| | Precision | Recall |
|---|---|---|
| DS0 | [20%, 20%, 20%, 20%, 20%] | [100%, 100%, 100%, 100%, 100%] |
| DS1 | [100%, 100%, 100%, 100%, 100%] | [100%, 100%, 100%, 100%, 100%] |
| DS2 | [100%, 100%, 100%, 100%, 100%] | [100%, 100%, 100%, 100%, 100%] |

From Fig. 3, it shows that TDEVs of DS0 are larger than those of DS1 and DS2. The reason is that some normal events are recognized as attacks, so the corresponding measured time offsets are not used for the update of the local clock. It can be shown from the analysis of precision and recall.

As shown in Table II, for the three algorithms, the precisions of the five paths are all 100%. The recalls of DS0 are 20%, and the recall of DS1 and DS2 are both 100%. Details analysis shows that, for each path 200 events are recognized as attack for DS0, and 40 TDAs are all detected for DS1 and DS2. For DS0, except for the 40 actual TDAs, the rest 150 events recognized as attack (false alarm) are all due to the attacks happened in other path. The simulation results prove that limiting maximum value of $m(attack)$ is an effective method to solve the problem of untrusted evidence due to the TDAs on other path when the path is normal.

The results of the first simulation is shown in Fig. 4. In the simulation, TDA happens once every 50s, and two paths are attacked each time, and the value of TDA is set to be 10ns. When $\mod(t_n, 50s) == 0$, TDAs happen for the 1'st and 2'nd path. When $\mod(t_n - 20s, 50s) == 0$, TDAs happen for the 3'rd and 4'th path. When $\mod(t_n - 40s, 50s) == 0$, TDAs happen for the 5'th path. For the simulation, 40 TDAs are applied to each path.

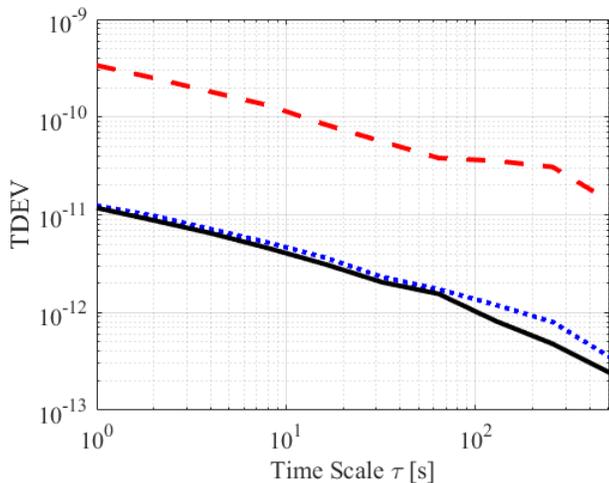

**Fig. 4.** Simulation of 5 paths time synchronization system's TDEV with two paths under simultaneous TDA once every 50s. (dot line(blue): DS0; dash line(red): DS1; solid line(black): DS2).

TABLE III
PERFORMANCE METRIC WITH ONE PATH UNDER TDA ONCE EVERY 50s

|  | Precision | Recall |
|---|---|---|
| DS0 | [33.3%, 33.3%, 33.3%, 33.3%, 33.3%] | [100%, 100%, 100%, 100%, 100%] |
| DS1 | [66.7%, 66.7%, 66.7%, 66.7%, 100%] | [75%, 75%, 82.5%, 82.5%, 100%] |
| DS2 | [100%, 100%, 100%, 100%, 100%] | [100%, 100%, 100%, 100%, 100%] |

From Fig. 4, it is shown that TDEVs of DS0 are larger than those of DS2, and TDEVs of DS1 are much larger than those of DS2 and DS0. The reason for DS0 is the same with the first simulation in Fig. 3, that some normal events are recognized as attacks, so the corresponding measured time offsets are not used for the update of the local clock. The reason for DS1 is that the introducing of maximum limitation of $m(attack)$ brings missed detections of TDAs which deteriorate the performance severely. The details analysis is shown below.

As shown in Table III, the precisions and recalls of the five paths are calculated for DS0, DS1, and DS2. Only for DS2, the precisions and recalls are both 100%. According to the definition of precision and recall, the number of false alarm and missed detection are both analyzed for DS0 and DS1. For DS0, 120 events are detected as attacks for each path, where 40 events are actual attacks and the rest 80 events are normal event. That means the number of false alarm is 80 and the number of missed detection is 0 for each path of DS0. For DS1, the events that detected as attacks are 45, 45, 40, 40, and 40 for the path 1, 2, 3, 4, and 5. The numbers of false alarm are 15, 15, 7, 7 and 0, respectively, and the numbers of missed detection are 10, 10, 7, 7, and 0, respectively. For the 5'th path, there are no false alarm and no missed detection since there is no simultaneous TDAs on any other path and the introduction of maximum limitation of $m(attack)$ solve the problem of untrusted evidence from other malicious path when the 5'th path is normal. For the 1'st, 2'nd, 3'rd, and 4'th path, the missed detection is not zero. On the one hand, the simultaneous TDAs on two paths is cancelled out when calculate the virtual time offset that brings wrong evidence with very high value of $m(normal)$. On the other hand, $m(attack)$ derived from other normal paths is limited by the maximum limitation. Then, the combination of the evidences brings wrong decisions. So, the attacks are not detected, which influence attack detection at following time and brings false alarms even when there is no attack. The simulation proves that limiting both maximum value and minimum value of $m(attack)$ is an effective method to solve the problem of untrusted evidence due to the TDAs on other path no matter whether the path is normal or malicious.

*B. Comparison of Secure Multiple Paths Combination Method with Other Methods*

In previous works, many methods have been proposed for multiple path synchronization. Especially, the FTA is thought as an efficient one to mitigate the path failure [15, 16]. In this subsection, we compare the performance of FTA and the D-S based method proposed in this paper. We also compare the performance between the attack detection method based on time offset of single path and the multiple paths methods. The algorithm proposed in [8] is adopted for attack detection method based on time offset of single path, which is denoted as Single method in the following.

In the simulation, 5 paths time synchronization system is studied to compare the performance of DS2 and FTA methods under four different conditions. For the Single method, only one path time synchronization system is simulated.

For the first condition, there is no TDA and no jump of slave clock. That means the five paths time synchronization system works in a normal condition. As shown in Fig. 5(a), the TDEVs of DS2 performs best. For DS2, all the time offsets are averaged to update the local clock. The TDEV of FTA method is larger than DS method, because the maximum and minimum time offsets are excluded even when there is no malicious path. The TDEVs of Single method are the largest one since there is only the time offset whose noise is larger than that of the five path system whose noises are averaged.

For the second condition, there is no jump of slave clock, however, TDAs happen in the 1'st and 2'nd paths simultaneously once every 50s. As shown in Fig. 5(b), the TDEVs of DS2 method is still the best. The precision and recall of the DS2 method is both 100%. Different from the first condition, the TDEV of FTA is even larger than the Single method. For FTA method, only the maximum and minimum time offset are excluded. However, there are two paths under simultaneous TDAs. So, TDA in one of the malicious path is not detected, and the wrong time offset from the malicious path is used to update the local clock, which deteriorate the performance of the synchronization system.

For the third condition, there is no TDA, however, the slave clock jump with 1ns happens once every 30s. As shown in Fig. 5(c), the TDEV of DS2 method is still the best one. No jump of slave clock is recognized as attack, and all the time offsets are averaged to update the slave clock. The difference between DS2 method and FTA method is due to the exclusion of the maximum and minimum time offset for FTA method. It is important to notice that even there is no attack, the TDEVs of Single method are much larger than that of the first condition. Details analysis shows that the jump of slave clock are recognized as attacks for Single method. So, the jump is not modified and it deteriorates the performance severely.

For the fourth condition, TDAs happen in the 1'st and 2'nd paths simultaneously once every 50s, and the slave clock jump with 1ns happens once every 30s. The TDAs and slave clock jump does not happened at the same time. As shown in Fig. 5(d), the curve of TDEVs of DS2 method is still the best one. The difference between the FTA method and DS2 method is the same with the second condition. The difference between the Single method and DS2 method is the same with the third condition.

In all, the DS2 shows the best performance no matter whether there are TDAs and whether there are slave clock jumps.

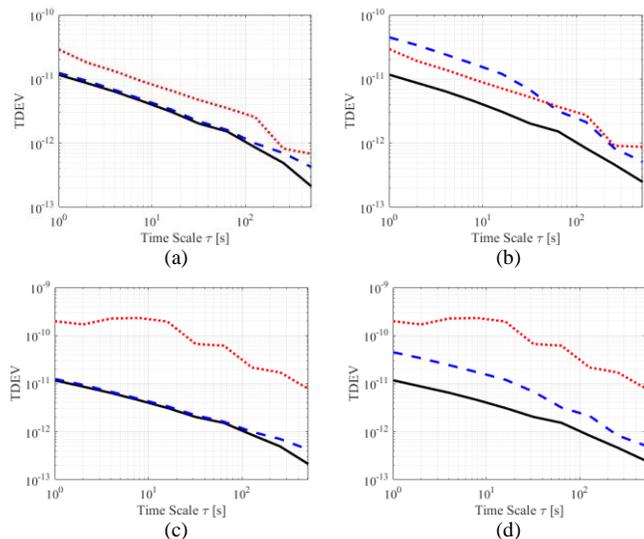

**Fig. 5.** Simulation of 5 paths time synchronization system's TDEV. (a) without TDA, without slave clock jump; (b) with two paths under simultaneous TDAs once every 50s, without slave clock jump ; (c) without TDA, with slave lock jump every 30s; (d) with two paths under simultaneous TDA once every 50s, with slave clock jump every 30s. (solid line(black): DS2; dash line(blue): FTA; dot line(red): Single).

## IV. EXPERIMENTAL DEMONSTRATION

In this section, in order to explore the feasibility of the proposed algorithms in this paper, experiments on three paths TWFTT system are demonstrated.

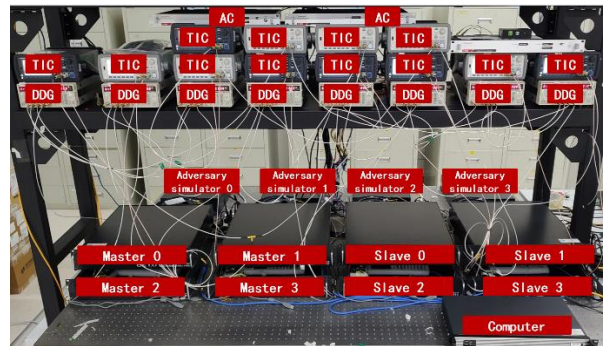

**Fig. 6.** Experimental setup of multiple paths TWFTT system with adversary simulators. AC: Atomic clock, TIC: Time Interval Counter, DDG: Digital Delay Generator.

As shown in Fig. 6, there are four TWFTT systems with adversary simulators, where three systems are used to test the proposed DS2 algorithms, and the rest one is used to initiate the system. Each TWFTT system is consist of one master part, one slave part, three time interval counters (TIC, Keysight 53230A), and two digital delay generators (DDG, SRS DG645). The master parties and slave parties of the TWFTT systems share common atomic clocks (AC), respectively.

For each TWFTT system, it is the same with the experimental setup in [8]. The wavelength division multiplexing scheme is adopted for the TWFTT system to confront the Rayleigh scattering in fiber. The channel 35/36 of Dense Wavelength Division Multiplexing (DWDM) are chosen for the optical signals from the master/slave site. So, the center wavelength of the CW lasers of master/slave site are set to be around 1549.32ns/1548.51ns. On the master/slave site, 1PPS optical signal is generated by modulating the light from the CW laser with an electro-optic modulator (EOM, AX-0S5-10-PFA-PFA-UL) which is driven by the 1PPS electric signal from DDG. The same 1PPS electric signal is sent to the start trigger port of the TIC at the same site. On each site, there is a photo detector, which transfers the 1PPS optical signal to 1PPS electric signal that is sent to the stop trigger port of the TIC on the same site. The master site sends the measured time interval ($\Delta T_M$) from the TIC in the master site to the slave site. Then $\theta_M = (\Delta T_M - \Delta T_S)/2$ is used as the measured time offset, where $\Delta T_S$ is the measured time interval from the TIC in the slave site. An extra TIC is used to measure the actual time offset between the master site and the slave site.

For each TWFTT system, an adversary simulator is added in fiber channel, which is used simulate the TDA launched by the adversary, similar to [24]. There is an optical switch in the adversary simulator. When the optical switch is set to one path, no extra asymmetry delay is added to the channel. When the optical switch is set to the other path, 2.5 ns asymmetry delay is added to the channel, which brings 1.25ns time error. In this paper, the TDA happens once every 50s. Besides that, 1ns is added to the slave DDG artificially to simulate slave clock jump every 30s.

In each round, the measured time offsets, $(\theta_M^1, \theta_M^2, \theta_M^3)$, from the three paths are sent to a computer, which runs the combination algorithm to detect the potential TDAs and calculate the update value of slave clock. The proposed DS2 method, the FTA method, and the Single method are all tested under the same condition. The TDEVs are show in Fig.6. For the Single method, it can detect and mitigate the TDA, but can not distinguish the slave clock jump from the TDA. For the FTA method, it can detect and mitigate the TDA and distinguish the slave clock jump from TDA. However, when the synchronization system is normal, it still use the information from one path for synchronization update. Only for the DS2 method, it can detect and mitigate the TDA and distinguish the slave clock jump from TDA. At the same time, it use the information from three paths for synchronization update. So, as show in Fig. 7, the DS2 method achieves better results than the other two methods. Detail analysis shows that all of the TDA are detected by the DS2 method, and all of the local clock jump is distinguished from the TDA. The time stability with 27.97 ps, 1.57 ps, and 1.12 ps at average time 1s, 10s, 100s is achieved with the DS2 method under TDA and local clock jump.

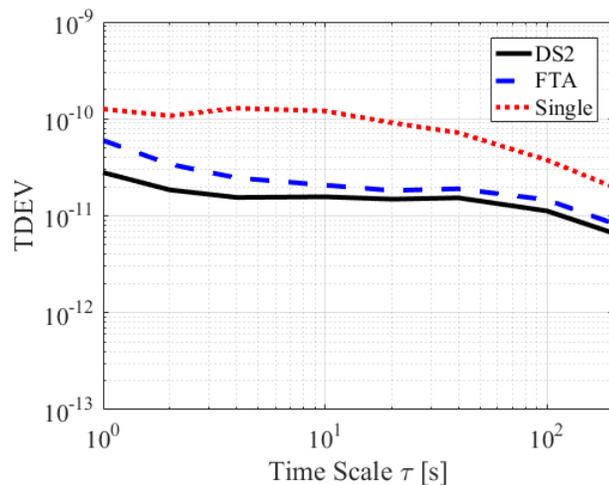

**Fig. 7.** Experimental demonstration of 3 paths time synchronization system's TDEV, with TDA once every 50s and slave lock jump every 30s. (solid line(black): DS2; dash line(blue): FTA; dot line(red): Single).

## V. DISCUSSION AND CONCLUSION

In this paper, we proposed a secure combination method based on the D-S theory for multiple paths synchronization. Two limitations are added to the basic probability assignment to optimize the combination method. Simulations show that the optimized method solves the problem of conflicting evidences. In order to verify the efficiency of the proposed method, a five paths time synchronization system is simulated under four different condition. It is shown that the proposed method shows better performance than FTA and Single method no matter whether there are TDAs or whether there are slave clock jumps, because the proposed method can not only detect TDA in the path but also distinguish slave clock jump from TDA. Furthermore, the feasibility and superiority of the proposed method is verified experimentally on 3 paths synchronization system The proposed secure combination method can be applied to improve the security and resilience of the synchronization protocols those are compatible with multiple paths architecture, such as NTP, PTP, and TWFTT.

ACKNOWLEDGMENT

This work was supported in part the Equipment Advance Research Field Foundation (Grant number 315067206), the National Key Research and Development Program of China (Grant No. 2020YFA0309704), the National Natural Science Foundation of China (Grant Nos U19A2076, 62101516, 62171418, 62201530), the Sichuan Science and Technology Program (Grant Nos 2022ZYD0118, 2023JDRC0017, 2023YFG0143, 2022YFG0330, 2022ZDZX0009 and 2021YJ0313), the Natural Science Foundation of Sichuan Province (Grant Nos 2023NSFSC1387 and 2023NSFSC0449), the Basic Research Program of China(Grant No. JCKY2021210B059), the Equipment Advance Research Field Foundation(Grant No. 315067206), the Chengdu Major Science and Technology Innovation Program (Grant No. 2021-YF08-00040-GX), the Chengdu Key Research and Development Support Program (Grant Nos 2021-YF05-02430-GX and 2021-YF09-00116-GX), the Foundation of Science and Technology on Communication Security Laboratory (Grant No. 614210304020121111).